\documentclass[%
 aip,
 jmp,%
 amsmath,amssymb,
 reprint,%
]{revtex4-1}

\usepackage{graphicx}
\usepackage{xcolor}

\usepackage[colorlinks=true,linkcolor=blue,anchorcolor=black,citecolor=black,filecolor=black,menucolor=black,runcolor=black,urlcolor=blue]{hyperref}

\newcommand{\hl}[1]{\textcolor{black}{#1}}

\begin{document}

\preprint{AIP/123-QED}

\title{The Entropic Barrier around the Conical Intersection Seam}

\author{Johannes C. B. Dietschreit}
\email{johannes.dietschreit@univie.ac.at}
\thanks{These authors contributed equally}%
\author{Sebastian Mai}%
 \email{sebastian.mai@univie.ac.at}
 \thanks{These authors contributed equally}%

\author{Leticia Gonz\'alez}
\affiliation{
 Institute of Theoretical Chemistry, Faculty of Chemistry, University of Vienna, Währinger Straße 17, 1090 Vienna, Austria
}%

\date{\today}

\begin{abstract}
Conical intersections (CIs) are seen as the main mediators of nonadiabatic transitions; yet, mixed quantum-classical (MQC) simulations that treat nuclei as classical point particles rarely, if ever, sample geometries with exactly degenerate electronic energies.  
Here we show that this behavior arises from a fundamental statistical–mechanical constraint on classical nuclear motion.  
Using a linear vibronic coupling model, we derive the free energy along the adiabatic energy gap and demonstrate analytically that as the gap approaches zero, the free energy diverges around the CI seam.  
Molecular dynamics simulations of the methaniminium cation on the S$_1$ surface confirm this prediction: trajectories can approach regions with small adiabatic gaps, but never reach the CI seam, even if the CI corresponds to a region of lowest potential energy.  
These results clarify why MQC methods successfully capture nonadiabatic behavior without sampling exact degeneracies and agree with recent findings that classical trajectories can sense the presence of CIs without visiting them.  

\end{abstract}


\maketitle

\section{Introduction}

Nonadiabatic transitions mediated by conical intersections (CIs) between adiabatic electronic surfaces represent a cornerstone of modern photochemistry and photophysics. 
Since the seminal work by Yarkony on “diabolical” intersections,\cite{Yarkony1996} and subsequent comprehensive reviews,\cite{Domcke2004,Domcke2012,Malhado2014} it has been recognized that CIs form \emph{seams} of degeneracy in nuclear configuration space, rather than isolated points, and that nuclear motion in their vicinity cannot be described reliably within the Born–Oppenheimer approximation. 
In a qualitative picture of excited-state dynamics, one often invokes a molecular wave packet transiting through a CI region, thereby enabling ultrafast internal conversion.

However, a full quantum mechanical description is cumbersome and limits simulations to very small systems. 
A solution is offered by mixed quantum--classical (MQC) methods, such as trajectory surface hopping (TSH), which treat nuclei classically and model the distribution of the nuclear wave packet through ensembles of trajectories.
Curiously, despite lacking a full nuclear wave function, and thus not explicitly encoding the geometric phase or the exact degeneracy manifold, MQC approaches frequently reproduce ultrafast nonadiabatic population transfer near CIs with reasonable accuracy (see, \textit{e.g.}, Gherib, Ryabinkin, and Izmaylov\cite{Gherib2015} or Malhado, Bearpark, and Hynes\cite{Malhado2014}). 
This strongly suggests that, in practice, what matters is not the nuclear coordinates reaching the exact degenerate geometry, but their passage through a region of strong nonadiabatic coupling around the CI seam.
Published distributions of the energy gap for those time steps where trajectories changed adiabatic surfaces in MQC simulations document that many hops occur far away from the CI seam; however, due to limited resolution, these histograms cannot exclude transitions occurring arbitrarily close to zero gap.\cite{Levine2007, Barbatti2021, Mukherjee2022, CoferShabica2022, Mansour2022, Mukherjee2024}
Even quantum-jump simulations based on Lindblad trajectories do not yield transitions at the exact degeneracy.\cite{Anderson2022}
It could even be shown that the CI does not have to be reached, but as long as the wave packet passes through regions with sufficiently strong coupling transfer between electronic states occurs.\cite{Farfan2019} 

A more quantitative perspective emerges from the statistical-mechanical analysis of Malhado and Hynes,\cite{Malhado2016} who used the Landau–Zener formula\cite{Landau1932,Zener1932} to derive a closed-form expression for nonadiabatic transition probabilities for a distribution of trajectories passing near a generic CI. 
They showed that the transition probability depends sensitively on the direction of approach and on the local CI topography in the branching plane.\cite{Domcke2012,Malhado2014} 
Furthermore, they demonstrated that trajectories undergoing state transitions pass near, but never exactly through, the CI.

\begin{figure}[tb]
    \centering
    \includegraphics[width=\linewidth]{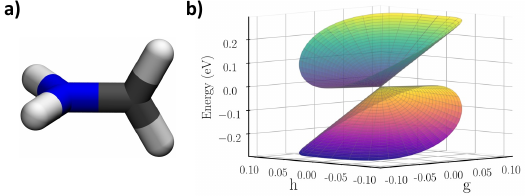}
    \caption{Conical intersection between the ground and first excited singlet states of the methaniminium cation (CH$_2$NH$_2^+$). 
    (a) Optimized geometry of the minimum-energy conical intersection (MECI) corresponding to the S$_1$/S$_0$ degeneracy point. 
    (b) Three dimensional rendering of the adiabatic potential-energy surfaces in the vicinity of the MECI, shown as cut along the branching-plane coordinates $g$ and $h$ (gradient-difference and nonadiabatic-coupling directions) to illustrate the characteristic double-cone topography of the CI.}
    \label{fig:CI_cut}
\end{figure}

In this work, we build on these insights and present a free-energy analysis showing that when nuclei are treated classically (as in most MQC schemes) and a collective variable (CV) is defined as the adiabatic energy gap, the equilibrium free energy at exactly zero gap is infinite. 
Thus, even though nonadiabatic transitions are enabled in the neighborhood of a CI, visiting the CI itself is statistically forbidden. 
We derive this result analytically for a generic linear vibronic-coupling model of a CI and discuss its implications for MQC simulations of electronically excited-state dynamics.
To complement the analytical treatment, we also perform explicit molecular dynamics simulations for the methanimine cation (CH$_2$NH$_2^+$, Fig.~\ref{fig:CI_cut}a) in its first excited singlet state. 
For this system, it is well established that the S$_1$/S$_0$ conical intersection corresponds to a point of lowest potential energy on the S$_1$ potential energy surface (PES).
To illustrate this, Fig.~\ref{fig:CI_cut} shows the minimum-energy conical intersection (MECI) of CH$_2$NH$_2^+$ and a visualization of the S$_1$ and S$_0$ potential-energy surfaces along the branching-plane coordinates.
Nevertheless, despite the apparent energetic driving force toward the MECI, and thus parts of the seam, our molecular dynamics simulations demonstrate that trajectories approach the seam but never reach the exact degeneracy, fully consistent with the statistical-mechanical argument developed below.

\section{Theory}

\subsection{Initial Definitions}

Following the derivation of Dietschreit \textit{et al.}\cite{Dietschreit2023}, the equilibrium free energy profile along a CV $\xi(\mathbf{R})$ is defined as
\begin{equation}
    F(z) = -k_\mathrm{B}T \ln \left[ \rho(z)\,\langle \lambda_\xi \rangle_z \right] ,
    \label{eq:free_energy}
\end{equation}
where $\mathbf{R} = (R_1, R_2, \ldots, R_{3N_a})^\top$ denotes a point in the nuclear configuration space of a system containing $N_a$ atoms.  
$\rho(z)$ is the marginal probability density of configurations satisfying $\xi(\mathbf{R}) = z$,
\begin{equation}
\begin{split}
    \rho(z) &= 
    \frac{\displaystyle \int \mathrm{d}\mathbf{R}\ 
    e^{-E(\mathbf{R})/k_\mathrm{B} T}\ 
    \delta\!\left(\xi(\mathbf{R})-z\right)}
    {\displaystyle \int \mathrm{d}\mathbf{R}\ 
    e^{-E(\mathbf{R})/k_\mathrm{B} T}} \\
    &= \left\langle \delta\!\left(\xi(\mathbf{R})-z\right) \right\rangle ,
\end{split}
\end{equation}
and $\lambda_\xi = \sqrt{h^2/(2\pi k_\mathrm{B}T\, m_\xi)}$ is the thermal de~Broglie wavelength of the CV, with $m_\xi$ the generalized mass in curvilinear coordinates.  
The notation $\langle \cdot \rangle_z$ denotes an ensemble average at fixed $\xi(\mathbf{R})=z$.

In this work, the CV is chosen as the energy gap between two consecutive adiabatic electronic states,  
\begin{equation}
    \xi(\mathbf{R}) = \Delta E_{IJ}(\mathbf{R}) 
    = E_{J}(\mathbf{R}) - E_{I}(\mathbf{R}) .
\end{equation}
The generalized CV mass in curvilinear coordinates is
\begin{equation}
 m_{\xi}^{-1} =  \sum_{a=1}^{N_a} \sum_{i=1}^{3} 
 \frac{1}{M_a} 
 \left(\frac{\partial \xi}{\partial R_{ai}} \right)^2 ,
\end{equation}
where $M_a$ is the mass of nucleus $a$.  
For the specific choice $\xi = E_{J} - E_{I}$ this becomes
\begin{equation}
 m_{\xi}^{-1}(\mathbf{R}) 
 = \sum_{a=1}^{N_a} \sum_{i=1}^{3} 
 \frac{1}{M_a} 
 \left( [\mathbf{F}_{I}]_{ai} - [\mathbf{F}_{J}]_{ai} \right)^2 ,
\end{equation}
where $\mathbf{F}_\alpha$ denotes the nuclear forces on state $\alpha$.  
If neither adiabatic potential diverges, the associated forces remain finite, implying that $m_\xi$ also stays finite.  
Because we focus on the behavior of the free energy near the CI and $\lambda_\xi$ remains finite, it suffices to analyze the limit of $\rho(z)$ as $z \to 0$.

\subsection{Approximating the Potential Energy Surface Near a Conical Intersection}

CIs are not isolated points but multidimensional hypersurfaces (seams) where two adiabatic states become degenerate, \textit{i.e.}, $\Delta E_{IJ}=0$.  
The degeneracy is lifted along two linearly independent directions in nuclear coordinate space, the gradient-difference and nonadiabatic-coupling directions, which span the branching plane.\cite{Yarkony1996,Yarkony2001,Yarkony2005}  
These are defined as
\begin{equation}
    \mathbf{g}_{IJ} = \frac{\partial}{\partial \mathbf{R}} (E_J - E_I) ,
    \label{eq:vector_g}
\end{equation}
and
\begin{equation}
    \mathbf{h}_{IJ} = (E_J - E_I)
        \left\langle \Psi_I^{\mathrm{el}} 
        \Big| \frac{\partial}{\partial \mathbf{R}} 
        \Big| \Psi_J^{\mathrm{el}} \right\rangle ,
         \label{eq:vector_h}
\end{equation}
where $\Psi_\alpha^{\mathrm{el}}$ are the adiabatic electronic wavefunctions.

Near a CI, the adiabatic potential energy matrix may be represented by a linear vibronic coupling model truncated after first-order terms,\cite{Atchity1991}
\begin{equation}
    \mathbf{V}(\mathbf{R}) =
    E_\mathrm{spec}(\mathbf{q}) +
    \begin{pmatrix}
        A g & C h \\
        C h & B g
    \end{pmatrix},
\end{equation}
where $g$ and $h$ are projections of $\mathbf{R}$ onto the branching vectors, $\mathbf{q}$ denotes the remaining $3N_a-8$ spectator modes, and $A$, $B$, and $C$ determine the local CI topography, namely its slope and tilt.  
Diagonalization gives the adiabatic energies
\begin{align}
    E_{J/I}(g,h,\mathbf{q}) & =
    E_{J/I}^\mathrm{CI}(g,h)+ E_\mathrm{spec}(\mathbf{q}) \nonumber \\
    & = \frac{A+B}{2}\, g
    \pm \frac{1}{2}\sqrt{
        \left(A-B\right)^2 g^2
        + 4C^2 h^2
    }
    + E_\mathrm{spec}(\mathbf{q}) .
    \label{eq:CI_energy_expression}
\end{align}

\subsection{Free Energy Near the Conical Intersection}

Using Eq.~(\ref{eq:CI_energy_expression}), the marginal Boltzmann distribution (the projection onto the energy gap) becomes
\begin{align}
    \rho_{J/I}(z)
    &= 
    \frac{
        \iiint \mathrm{d}g\,\mathrm{d}h\,\mathrm{d}\mathbf{q}\,
        e^{-E_{J/I}(g,h,\mathbf{q})/k_\mathrm{B}T}
        \delta(\Delta E - z)
    }{
        \iiint \mathrm{d}g\,\mathrm{d}h\,\mathrm{d}\mathbf{q}\,
        e^{-E_{J/I}(g,h,\mathbf{q})/k_\mathrm{B}T}
    } \nonumber \\
    & = \frac{
            \iint  \mathrm{d}g\ \mathrm{d}h\ e^{-E_{J/I}^\mathrm{CI}(g,h)/k_\mathrm{B}T}\ \delta(\Delta E - z)
        }{
            \iint  \mathrm{d}g\ \mathrm{d}h\ e^{-E_{J/I}^\mathrm{CI}(g,h)/k_\mathrm{B}T} 
        } \ .
    \label{eq:rhoz_gh}
\end{align}
Assuming orthogonality between branching and spectator coordinates near the CI allows factorization of the $\mathbf{q}$ integrals, which cancel in numerator and denominator.

Recognizing that the square root term in Eq.~(\ref{eq:CI_energy_expression}) equals $\Delta E$, we introduce the coordinate transformation
\begin{align}
    \Delta E &= \sqrt{\left(A-B\right)^2\, g^2 + 4C^2 h^2}, \\
    \phi &= \mathrm{atan2}\!\left( C h, \tfrac{A-B}{2} g \right),
\end{align}
which carries Jacobian
\begin{equation}
\mathbf{J} = \left( 
    \begin{array}{cc}
        \frac{\partial g}{\partial \Delta E} & \frac{\partial g}{\partial \phi} \\
        \frac{\partial h}{\partial \Delta E} & \frac{\partial h}{\partial \phi} 
    \end{array}
    \right)
    = 
    \left( 
    \begin{array}{cc}
        \frac{\cos \phi}{A-B} & -\frac{\Delta E \sin \phi}{A-B} \\
        \frac{\sin \phi}{2C} & \frac{\Delta E \cos \phi}{2C} 
    \end{array}
    \right)
\end{equation}
and
\begin{equation}
    |\mathbf{J}| = 
    \frac{\Delta E}{2|(A-B)C|} \ .
    \label{eq:abs_jacobian}
\end{equation}
In these coordinates,
\begin{equation}
    E_{J/I}(\Delta E,\phi)
    = \frac{1}{2} \frac{A+B}{A-B}\,
        \Delta E \cos\phi
      \pm \frac{\Delta E}{2} \ .
\end{equation}

Substituting this into Eq.~(\ref{eq:rhoz_gh}) yields
\begin{align}
    \rho_{J/I}(z) 
    & \propto \int_0^{2\pi} \mathrm{d}\phi \int_0^\infty \mathrm{d}\Delta E\ |\mathbf{J}|\ e^{-E(\Delta E, \phi)/k_\mathrm{B}T}\ \delta(\Delta E - z) \nonumber \\
    & \propto 
    \int_0^{2\pi}\!\mathrm{d}\phi\
    \frac{z}{2(A-B)C}\, \nonumber \\
    & \times \exp\!\left[
        -\frac{1}{k_\mathrm{B}T}
        \left(
           \frac{1}{2}
            \frac{A+B}{A-B}
            z \cos\phi 
            \pm \frac{z}{2}
        \right)
    \right]\ .
\end{align}
The angular integral is a modified Bessel function of the first kind, $I_0(\Gamma)$, with $\Gamma(z) = \tfrac{1}{2}\tfrac{A+B}{A-B}\tfrac{z}{k_\mathrm{B}T}$, giving
\begin{equation}
    \rho_{J/I}(z) 
    \propto 
    \frac{\pi}{(A-B)C}\,
    \times e^{\mp z/(2k_\mathrm{B}T)}\,
    \times I_0\!\left(\Gamma(z)\right) \times  \ z.
    \label{eq:dens_near_CI}
\end{equation}
Both the exponential and $I_0(\Gamma)$ approach unity as $z=\Delta E \to 0$, so that
\begin{equation}
    \lim_{\Delta E \to 0} \rho_{J/I}(\Delta E) = 0 \ ,
    \label{eq:limit_dens_CI}
\end{equation}
agreeing with the previous finding that the CI seam is a zero measure manifold\cite{Malhado2016}.
Since $\langle \lambda_\xi \rangle_z$ remains finite as $\Delta E\to0$, the free energy for either electronic state  diverges at the CI,
\begin{align}
    \lim_{\Delta E\to0} F_{J/I}(\Delta E)
    & = \lim_{\Delta E\to0} \Bigl[ & -k_\mathrm{B}T \ln (\Delta E) \mp \frac{\Delta E}{2}  \nonumber \\
    && - \frac{(A+B)^2}{32(A-B)^{2}}\frac{(\Delta E)^2}{(k_\mathrm{B}T)^2} \nonumber \\
    && + O((\Delta E)^4) + \mathrm{const.} \Bigl] \nonumber \\
    & = \infty,
    \label{eq:freeenergy_limit}
\end{align}
for any nonzero temperature $T$, where the first line is a Taylor expansion of the free energy around zero.  
Thus, independent of the actual topography of the CI and even if it lies at the minimum of an adiabatic PES, the free-energy creates an insurmountable retaining force, preventing a classical  trajectory from reaching the exact intersection seam.

\subsection{Entropic Contribution}

To better understand the behavior of the free energy, we inspect the internal energy $U$ and entropy $S$. 
The internal energy profile is defined as\cite{Dietschreit2023}
\begin{equation}
    U_{J/I}(\Delta E) = 
    \frac{\left\langle E_{J/I}\, m_\xi^{-1/2} \right\rangle_{\Delta E}}
         {\left\langle m_\xi^{-1/2} \right\rangle_{\Delta E}} .
\end{equation}
Because both the adiabatic energies and the CV mass remain finite for all $(g,h)$, the internal energy does not diverge near the CI.  
Consequently, the divergence of the free energy must be entirely entropic.

Under equilibrium conditions, the entropy of the canonical ensemble is conjugate to the free energy and has a thermodynamic interpretation:
\begin{equation}
    S(\Delta E) = -\frac{\partial F(\Delta E)}{\partial T}
          = \frac{U(\Delta E) - F(\Delta E)}{T}\ .
          \label{eq:S_canonical}
\end{equation}

As a more general measure of configurational entropy that can be applied to any well-defined distribution of nuclear configurations, we compute the energy-gap conditioned Shannon entropy as:
\begin{align}
    S_\mathrm{Shannon}(\Delta E) & = - k_\mathrm{B}\, \langle \ln P\rangle_{\Delta E} \\
    & = - k_\mathrm{B} \int \mathrm{d}\mathbf{R}\ P(\mathbf{R} | \Delta E)\, \ln P(\mathbf{R} | \Delta E) \ ,
    \label{eq:S_Shannon}
\end{align}
where $P(\mathbf{R} | \Delta E)$ is the full-dimensional configurational probability distribution, conditioned on the energy gap $\Delta E$.

We expect both entropy measures to diverge as $\Delta E \to 0$.
Even though the onset of the divergence might occur for slightly different values of $\Delta E$, it has the same origin: the distribution $P$ or probability density $\rho$ loses support on the shrinking $(g,h)$ branching-plane shell, so that $S \to -\infty$ logarithmically.
This is similar to the particle-in-a-box, where the energy levels diverge as the accessible volume shrinks to zero, even though the exact scaling is different.

\subsection{Geometric Origin}
\label{sec:geometric_origin}

Even though the derivation of Eq.~(\ref{eq:limit_dens_CI}) was carried out under equilibrium conditions, its implications hold beyond the canonical ensemble.
Due to the $\Delta E$ term in the Jacobian determinant in Eq.~(\ref{eq:abs_jacobian}) any marginal probability density $p(\Delta E)$ obtained from a smooth, non-singular density $P(g,h,\mathbf{q})$ will satisfy near the CI seam
\begin{equation}
    p(\Delta E) \propto \frac{\Delta E}{2|(A-B)C|} \ .
\end{equation}
Since the CI seam is a $3N_a-8$ submanifold, this statement is independent of whether $P(g,h,\mathbf{q})$ is a equilibrium Boltzmann distribution, from a swarm of TSH trajectories, or produced by any other arbitrary non-equilibrium process, as long as it does not diverge at $\Delta E =0$.

\hl{%
However, this geometric statement is conditional on the prefactor of the marginal density being non-singular. 
For canonical sampling this follows from the Boltzmann weight; for non-equilibrium dynamics we do not have a general proof. 
Therefore, for general intersections and arbitrary non-equilibrium initial conditions, the statement that classical trajectories never access the CI seam has to remain a conjecture rather than a proven theorem.
}

\section{Computational Details}

All excited-state molecular dynamics simulations were performed using SHARC~4.0\cite{SHARC4} and the electronic structure calculations with \textsc{OpenMolcas}~v24.06.\cite{FdezGalvn2019, Aquilante2020, LiManni2023}
Specifically, the methaniminium cation (CH$_2$NH$_2^+$) was propagated at the state-averaged complete active space self-consistent field (SA-CASSCF) level of theory, averaging over two singlet states with a CAS(2,2) active space consisting of the $\pi$ and $\pi^*$ orbitals and using the cc-pVDZ basis set.\cite{Roos1980, Werner1985, Andersson1992, Dunning1989}  
This minimal active space was chosen instead of the more common CAS(6,4),\cite{Westermayr2019} as it provides superior numerical stability for our purposes and higher excited states are not relevant for the present study.  
The Cholesky decomposition\cite{Pedersen2009} was employed throughout the CASSCF calculations, and analytical gradients and nonadiabatic coupling vectors were computed at every step.
We optimized CH$_2$NH$_2^+$ in the ground state and performed a frequency calculation.

\subsection{Adiabatic Dynamics}

For the adiabatic excited state simulations, we sampled 200 phase-space points from the Wigner distribution.
All initial conditions were then manually initialized in the first excited singlet state (S$_1$).

We performed adiabatic dynamics in the S$_1$ state using the velocity--Verlet algorithm with a time step of 0.5\,fs.
Each trajectory was simulated for 5000\,fs.
Thermal effects were modeled using a Langevin thermostat set to 1000~K with a friction coefficient of 0.01~fs$^{-1}$ to sample the canonical ensemble.
For adiabatic dynamics, surface hopping was disabled; consequently, the details of electronic propagation, decoherence correction, and kinetic energy adjustment did not affect the dynamics.
Gradients and nonadiabatic coupling vectors of both states were computed and saved at every time step for subsequent analysis.

The elevated temperature accounts for the conversion of potential into kinetic energy stemming from the relaxation from the Franck-Condon region towards the CI seam, which significantly heats up the small molecule.
Furthermore, the simulation conditions represent a stress test for our predictions; the simulation time allows the system to spend a long time near the CI seam, and at 1000~K, a failure to reach the seam cannot be attributed to insufficient thermal energy.

To remove nonequilibrium effects from the vertical excitation, the first 250\,fs of each trajectory were discarded.  
Four of the 200 trajectories terminated prematurely due to CASSCF convergence failures at very high internal energies and were excluded entirely.  
All remaining data points were included in the analysis presented in this work.  
Free-energy profiles were computed using tools from the \texttt{adaptive\_sampling} package.\cite{Hulm2022, ochsenfeldadaptive_sampling}

In addition to the adiabatic energy gap, we computed the minimal root-mean-squared distance (RMSD) between each instantaneous geometry and the minimum-energy conical intersection (MECI) structure.  
We also evaluated scalar projections of each geometry onto the unit gradient-difference vector $\mathbf{g}$ (Eq.~(\ref{eq:vector_g})) and the nonadiabatic-coupling vector $\mathbf{h}$ (Eq.~(\ref{eq:vector_h})).  
Geometry alignment using the Kabsch algorithm\cite{Kabsch1976, Kabsch1978} and subsequent RMSD and projection calculations were implemented in \textsc{PyTorch}\cite{Pytorch2019} to enable access to the Cartesian gradient of the reaction coordinates.  
Because two symmetry-equivalent MECI geometries exist due to the permutation of the hydrogen atoms (corresponding to rotating either the CH$_2$ or NH$_2$ fragment by 180$^\circ$ around the C–N bond), both were considered when computing RMSD and branching-plane projections. 
We always selected the MECI geometry that had the smaller RMSD after alignment.

\subsection{Surface Hopping Dynamics}
\label{sec:methods-fssh}

In addition to the thermostatted, canonical dynamics constrained to S$_1$, we also performed fewest switches surface hopping (FSSH) to generate a nonadiabatic ensemble.
For the FSSH dynamics, we used the same level of theory and number of states as for the adiabatic excited state dynamics.
We sampled 5000 phase-space points from the Wigner distribution and initialized them in the S$_1$ state.
The nuclei were propagated with the velocity-Verlet integrator using a time step of 0.5\,fs and 25 electronic substeps; no thermostat was applied. 
The electronic wave function was propagated with the local diabatization method\cite{Granucci2001} using overlaps computed within OpenMolcas with the restricted active space state interaction module.\cite{Malmqvist1989}
Each trajectory was propagated for 100\,fs irrespective of the active electronic state, so that frames following downward hops are included. 
The energy-based decoherence correction of Granucci and Persico\cite{Granucci2007} was applied with $C=0.1$\,Ha.
At successful hops, nuclear velocities were rescaled along the nonadiabatic coupling vector to conserve total energy; at frustrated hops, the velocity component along the nonadiabatic coupling vector was reversed.
Of the 5000 trajectories initiated, 93 were discarded due to CASSCF convergence failures.

\subsection{Configurational Shannon Entropy conditioned on the energy gap}
\label{sec:shannon-method}

The configurational Shannon entropy conditioned on $\Delta E$  (Eq.~(\ref{eq:S_Shannon})) was computed for three ensembles using an identical protocol: (i)~the canonical $\mathrm{S}_1$ trajectories at $T=1000$~K, (ii)~the full ensemble of FSSH trajectories pooled over all frames irrespective of the active state, and (iii)~the subset of FSSH frames preceding the first downward hop, which constitutes a non-equilibrium excited-state ensemble unaffected by surface hopping.

Each frame was mapped onto a translation- and rotation-invariant descriptor consisting of all pairwise interatomic distances. 
To control the dimensionality, principal component analysis was performed on the pooled descriptor vectors from the canonical and FSSH simulations, and the leading ten components were retained, accounting for $98\%$ of the cumulative variance. 
The same orthogonal basis was applied to all three ensembles so that their entropies are evaluated on a common set of configurational coordinates. 
A one-dimensional histogram with $100$ bins spanning the range of the pooled projected data was constructed for each principal component, and the $\Delta E$ grid was likewise fixed prior to the analysis. 
Within every $\Delta E$ bin containing at least ten samples, the entropy was computed for each component and summed,
\begin{equation}
 S_\mathrm{Shannon}(\Delta E) / k_\mathrm{B} \approx
    -\sum_{k=1}^{10}\sum_{i=1}^{100} P(k,i|\Delta E) \, \ln P(k,i|\Delta E) \ ,
  \label{eq:shannon-est}
\end{equation}
where $P(k,i|\Delta E)$ is the fraction of samples in the $\Delta E$ bin falling into the $i$-th histogram bin of principal component $k$. 

Equation~\eqref{eq:shannon-est} corresponds to assuming statistical independence of the principal components within each $\Delta E$ bin. 
This approximation is necessary, since a fully joint histogram with $100^{10}$ bins cannot easily be realized.
Because all three ensembles are processed in the same PCA basis, on the same per-component grids, the differences between their configurational entropies remain comparable.

\section{Results and Discussion}

\subsection{Excited State Equilibrium Dynamics}

\begin{figure}[tb]
    \centering
    \includegraphics[width=\linewidth]{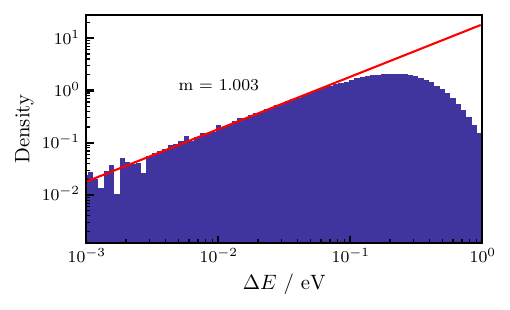}
    \caption{Histogram of the adiabatic energy gap $\Delta E$ between the ground and first excited singlet states from the canonical excited state simulations. 
    In the log–log representation, the low-$\Delta E$ region exhibits a power-law behavior; the fitted slope $m$ corresponds to the polynomial exponent governing the dependence of the density on $\Delta E$.}
    \label{fig:gap_histogram}
\end{figure}

The log--log histogram of the sampled adiabatic energy gap $\Delta E$ (Fig.~\ref{fig:gap_histogram}) shows a clear linear decrease in population as the gap approaches zero.  
This behavior is fully consistent with the theoretical prediction of Eq.~(\ref{eq:dens_near_CI}), which also yields a linear dependence of the marginal probability density $\rho(\Delta E)$ for small gaps.  
Thus, the adiabatic molecular dynamics trajectories exhibit precisely the scaling expected for a system in which the accessible configuration-space volume shrinks near a CI seam.

\begin{figure}[tb]
    \centering
    \includegraphics[width=\linewidth]{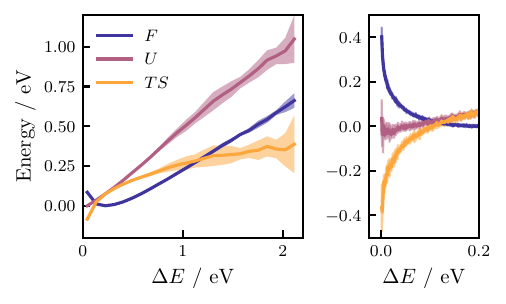}
    \caption{Free energy $F$, internal energy $U$, and entropy $T S$ profiles from the S$_1$ dynamics of CH$_2$NH$_2^+$ as a function of the adiabatic energy gap $\Delta E$. 
    The right panel shows a magnified view of the small-gap region. 
    The minimum of the free-energy curve as well as $U(0)$ were shifted to zero for clarity.
    All energies are given in electron volts.
    The shaded regions indicate the standard deviation between curves obtained by splitting the 196 trajectories into batches of 28 each.
    The solid line represents the average of the seven curves computed this way.}
    \label{fig:profiles_deltaE}
\end{figure}

The free-energy analysis is shown in Fig.~\ref{fig:profiles_deltaE}.
The internal-energy profile  $U$ decreases almost perfectly linearly with the energy gap over a broad range of $\Delta E$ values.  
The fluctuations at larger gaps arise from limited sampling, as trajectories rapidly leave the Franck--Condon region and move toward the S$_1$/S$_0$ conical-intersection seam. 
As anticipated from the analytical model, the internal energy remains finite throughout: the S$_1$ potential-energy surface is smooth in all directions except at the degeneracy itself, and the generalized CV mass is likewise finite.   
Consequently, there is no mechanism by which the internal energy $U$ could diverge as $\Delta E \to 0$.

In contrast, the free-energy profile exhibits a more complex behavior.  
For most gap values, it decreases approximately linearly, mirroring the trend in $U(\Delta E)$.  
However, near $\Delta E \approx 0.2$\,eV the free energy develops a local minimum, followed by a rapid increase as the gap becomes smaller.  
This increase reflects precisely the divergence predicted by the linear vibronic-coupling model in Eq.~(\ref{eq:freeenergy_limit}): as the energy gap approaches zero, the entropic penalty associated with the vanishing phase-space volume dominates the free-energy expression, driving $F(\Delta E)$ upward.

The entropic contribution is shown explicitly in the entropy profile, plotted as $T S$ in Fig.~\ref{fig:profiles_deltaE}.  
The entropy decreases monotonically with decreasing energy gap, consistent with a progressive reduction of the accessible phase-space volume in the branching-plane coordinates.  
Near $\Delta E = 0$, the decay becomes increasingly steep: the number of accessible microstates $W(\Delta E) = |\{\mathbf{R}: \xi(\mathbf{R}) = \Delta E \}|$ collapses to zero, yielding a diverging Boltzmann entropy
\begin{equation}
    \lim_{\Delta E \rightarrow 0} S 
    = \lim_{W\rightarrow 0} k_{\mathrm{B}} \ln W 
    = -\infty .
\end{equation}
This insurmountable barrier dominating the free-energy expression is purely entropic in nature, as it is associated with the vanishing configuration-space volume and not with a rise in potential energy.
The simulations thus confirm the central theoretical prediction: the free energy contains an entropic contribution that prevents classical nuclear trajectories from sampling configurations exactly on the CI seam.

The combined simulation and analytical results, therefore, provide a coherent picture.  
Although the S$_1$/S$_0$ MECI in methaniminium corresponds to an energetically favorable region (\textit{i.e.}, it corresponds to the configuration of lowest energy on the S$_1$ surface), the free-energy associated with the energy-gap CV diverges.  
As a consequence, the CI is never visited in practice, even in the long MQC trajectories propagated at elevated temperatures presented here.  
Trajectories may approach the seam arbitrarily closely, but the exact degeneracy manifold remains statistically inaccessible.

\begin{figure}[tb]
    \centering
    \includegraphics[scale=1]{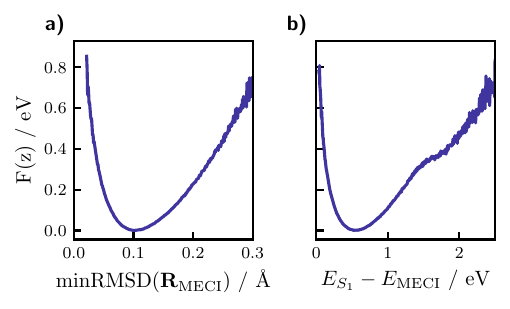}
    \caption{Free-energy profile $F(z)$ as a function of a) the minimal RMSD from the MECI geometry and b) the difference between the S$_1$ energy and the S$_1$ energy at the MECI.}
    \label{fig:rmsd_relE}
\end{figure}

To view the same effect from a different perspective, we also computed free-energy profiles according to Eq.~(\ref{eq:free_energy}) for two additional CVs:  
(i) the minimal root-mean-squared deviation (RMSD) from the MECI geometry, and  
(ii) the difference between the instantaneous S$_1$ energy and the S$_1$ energy of the MECI.  
Both CVs quantify the distance from the MECI geometry, but do not measure the distance to higher-energy points along the CI seam.  
Nevertheless, the free-energy profiles shown in Fig.~\ref{fig:rmsd_relE} diverge as the CV approaches zero, demonstrating that the MECI geometry itself is never visited --- even though it would be the most attractive point on the S$_1$ surface when potential energy alone is considered.

Another way to characterize the deviation from the CI is to project the Cartesian displacement vector of a configuration from the MECI onto the branching plane, which is spanned by the gradient-difference vector $\mathbf{g}$ (Eq.~(\ref{eq:vector_g})) and the nonadiabatic-coupling vector $\mathbf{h}$ (Eq.~(\ref{eq:vector_h})) of the MECI geometry.  
The resulting scalar coordinates indicate how far a geometry lies from the MECI along each branching-plane direction; the MECI is reached only when both $g$ and $h$ are simultaneously zero.

\begin{figure}[tb]
    \centering
    \includegraphics[scale=1]{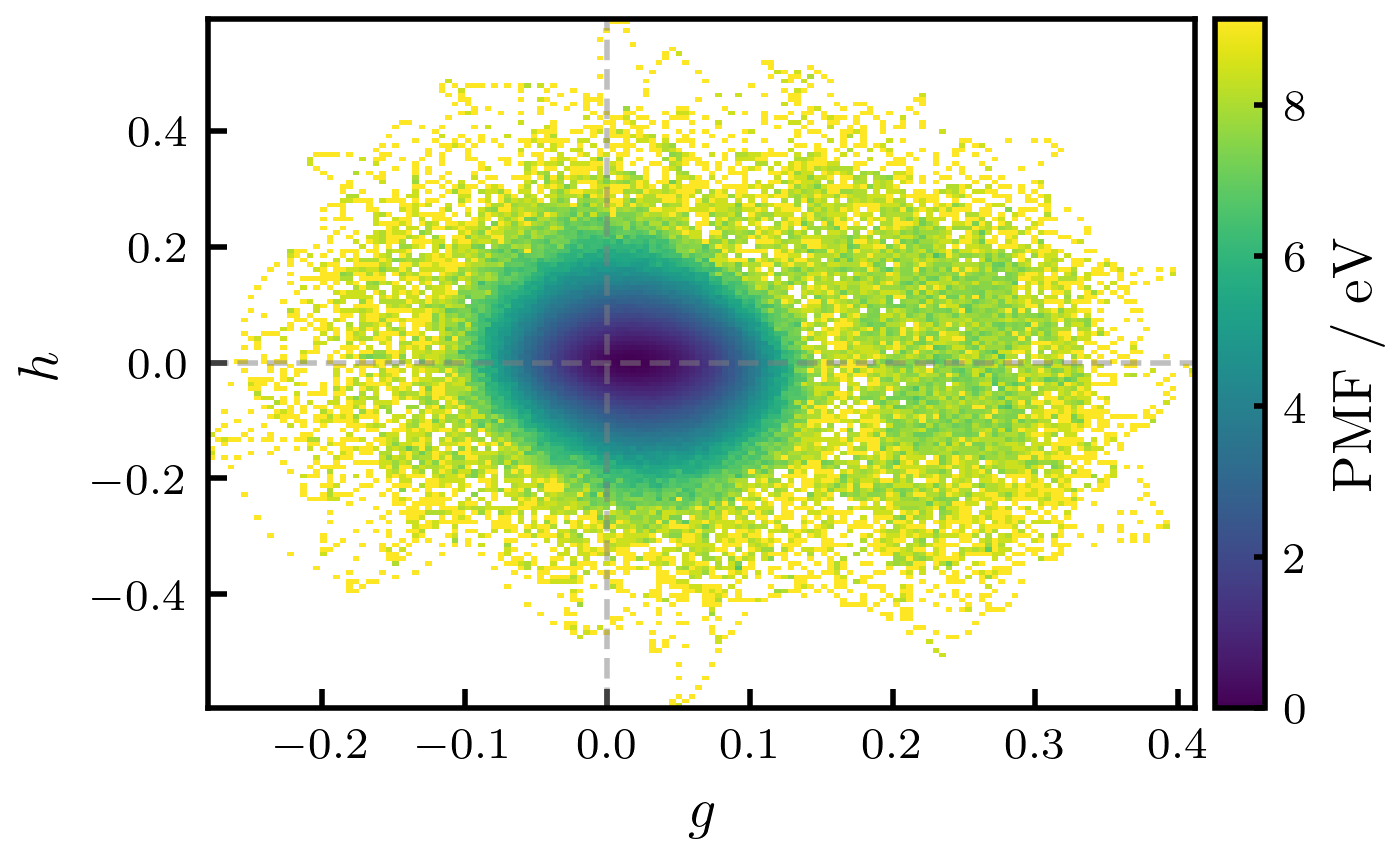}
    \caption{Two-dimensional potential of mean force (PMF) along the projections onto the gradient-difference coordinate $g$ and the nonadiabatic-coupling coordinate $h$. 
    The dashed gray lines mark $g = 0$ and $h = 0$, whose intersection corresponds to the MECI geometry.  
    }
    \label{fig:projection_g_h}
\end{figure}

The potential of mean force (PMF), defined in two dimensions as
\begin{equation}
    A(g, h) = -k_\mathrm{B}T \ln \left[ \rho(g,h) \right] 
\end{equation}
with $\rho(g,h) $ being the marginal Boltzmann probability density along the projected $g$ and $h$ values, is shown in Fig.~\ref{fig:projection_g_h}.  
Contrary to what one might expect from the previous results, the PMF exhibits a shallow minimum slightly displaced from the origin; however, at the origin, no divergence is visible.  
The reason for this apparent discrepancy is the different dimensionality of this projection. 
In the one dimensional representations, such as $\Delta E$, we have effectively transformed the projection onto the $g-h$-plane into polar coordinates and integrated over the polar angle. 
The configuration space represented by the circle with a radius of zero (the CI seam) has no volume and thus carries no statistical weight. 
More rigorously, lower-dimensional manifolds such as the CI seam, which is a $(3N_a -8)$-dimensional space, have zero measure under continuous distributions.
This means that the probability for a trajectory to pass arbitrarily close by, however, is finite. 

\subsection{Non-equilibrium nonadiabatic dynamics}

The analysis above relies on canonical, adiabatic sampling of the excited state, which excludes nonadiabatic transitions by construction. 
To verify that the linear depletion of the marginal distribution along the energy gap near the seam extends beyond the canonical ensemble, we performed FSSH simulations of the same system, drawing initial conditions from the same ground-state Wigner distribution and using identical electronic structure (see Sec.~\ref{sec:methods-fssh}).

\begin{figure}[tb]
    \centering
    \includegraphics[scale=1]{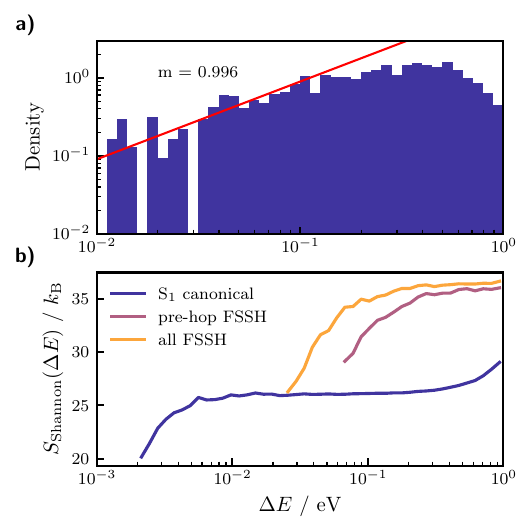}
    \caption{%
    a) Distribution of the adiabatic $\mathrm{S}_1$--$\mathrm{S}_0$ energy gap $\Delta E$ at the first downward hop in the FSSH simulations.
    In the double-logarithmic representation, the low-$\Delta E$ region is consistent with a power-law form $p(\Delta E) \propto \Delta E^{m}$;
    the fitted exponent $m$ is indicated. 
    b) Configurational Shannon entropy conditioned on the energy gap, Eq.~(\ref{eq:S_Shannon}), for the canonical excited-state simulations (blue), the FSSH
    trajectory segments preceding the first downward hop (magenta), and the full FSSH ensemble (yellow). 
    $S_\mathrm{Shannon}$ is different from the canonical $T\cdot S$ shown in Fig.~\ref{fig:profiles_deltaE}.
    }
    \label{fig:fssh_entropy}
\end{figure}

Figure~\ref{fig:fssh_entropy}a shows the distribution of energy gaps at the moment of the first downward hop, a quantity commonly reported in the surface-hopping literature. 
Despite 4907 successful trajectories, this distribution is sampled by a single frame per trajectory and is correspondingly noisier than the histogram of Fig.~\ref{fig:gap_histogram}. 
A linear least-squares fit to the double-logarithmic histogram over the range $0.01$--$0.1$\,eV nevertheless recovers a slope of approximately one, consistent with the linear scaling of $p(\Delta E)$ predicted by the geometric argument also at hop events. 
The smallest gap realized across the ensemble is $\Delta E_\mathrm{min} = \,$0.0025\,eV, confirming that no trajectory reaches the seam.

The canonical entropy of Eq.~\eqref{eq:S_canonical}  is not defined for non-equilibrium dynamics. 
To compare ensembles on the same footing, we therefore evaluated the configurational Shannon entropy conditioned on the energy gap, Eq.~(\ref{eq:S_Shannon}), in three cases: 
the thermostatted adiabatic excited-state ensemble, the subset of FSSH frames preceding the first downward hop, a non-equilibrium $\mathrm{S}_1$ ensemble unaffected by surface hopping, and the full FSSH ensemble pooled over all frames irrespective of the active state. 
Figure~\ref{fig:fssh_entropy}b shows that all three $S_\mathrm{Shannon}(\Delta E)$ curves decrease sharply as $\Delta E \to 0$. 
It is interesting to note that the full FSSH ensemble and the pre-hop ensemble show almost identical curves, whereas the canonical excited state simulations produce a markedly different configurational entropy curve.
The value of $\Delta E$ at which each curve ends is set by the smallest gap accessed in the corresponding ensemble and reflects the finite resolution of the conditional histogram.

The collapse of the configurational entropy near the seam is therefore not specific to canonical sampling: 
it persists for FSSH trajectories that have not yet undergone any nonadiabatic transition and for the full ensemble including post-hop dynamics on $\mathrm{S}_0$. 
This confirms that the entropic suppression of seam access is a generic geometric feature of the configuration space near a conical intersection, and does not depend on the dynamics used to sample it.

\section{Conclusion}

We have presented both analytical and simulation-based evidence that, when nuclei are treated classically, they are barred from accessing the conical-intersection (CI) seam by an entropic barrier due to the collapse of available configurational space.
As discussed in Section~\ref{sec:geometric_origin}, this behavior is geometric in nature and analogous to the vanishing probability of zero relative distance in radial distribution functions or zero velocity in Maxwell–Boltzmann statistics, where geometry alone suppresses probability density even in the absence of an energetic barrier.

The gap-based free-energy framework developed here thus identifies the CI seam as effectively “forbidden territory” for classical trajectories, even though its topography usually brings trajectories into its vicinity.
The linear vibronic coupling model predicts that the marginal probability density vanishes as $\Delta E \to 0$, and our molecular dynamics simulations of methaniminium (CH$_2$NH$_2^+$) confirm that trajectories can approach arbitrarily small gaps but never reach the exact degeneracy.

This is important as algorithms like metaFALCON\cite{Lindner2019metafalcon} that try to discover and sample CI seams using enhanced sampling will only get close to the CI seam, but never discover configurations that actually lie on it. 
Zero gap configurations will only be obtained through some constrained geometry optimization and not dynamics.

Our findings complement the recent observations of Karmakar, Thakur, and Jain,\cite{Karmakar2024} who concluded that purely classical trajectories can \emph{sense} the presence of a CI without ever arriving at it.  
Taken together, these results reinforce the notion that although the CI seam exerts a strong influence on nonadiabatic dynamics, it remains statistically inaccessible by trajectories evolving on adiabatic surfaces with classical nuclear motion.

From the perspective of computational dynamics, our findings underscore that MQC simulations that treat nuclei classically rely on sampling the vicinity of the CI, where nonadiabatic couplings are large, rather than populating the degeneracy manifold itself.  
This has direct implications for interpreting reaction pathways, transition-state analogies in excited states, and designing enhanced-sampling strategies for nonadiabatic processes.

\begin{acknowledgments}
The authors thank Carolin Müller for the code to visualize the CI. 
Further, the Austrian Scientific Cluster (VSC5) is acknowledged for the generous allocation of computational resources and the University of Vienna for continuous support.
We also thank Christoph Dellago for his groundbreaking contributions to the fields of statistical mechanics and molecular simulations, which continue to inspire us.
\end{acknowledgments}

\section*{Data Availability Statement}
All simulation data and analysis scripts have been deposited in the Zenodo archive at \url{https://doi.org/10.5281/zenodo.18196244}.
\hl{Additionally, the trajectories have been added to the SHNITSEL (Surface Hopping Nested Instances Training Set for Excited-state Learning) repository at \url{https://doi.org/10.5281/zenodo.14910194} to facilitate the development of machine learning potentials.}

\bibliography{references}

\end{document}